\documentclass[sigconf,screen]{acmart}
\makeatletter
\global\@ACM@balancefalse
\makeatother

\usepackage{tikz}
\usetikzlibrary{positioning, shapes.geometric, arrows.meta, calc, decorations.pathmorphing, shadows, backgrounds}

\definecolor{serverColor}{RGB}{74, 144, 226}          %
\definecolor{safeColor}{RGB}{34, 139, 34}             %
\definecolor{intermediateColor}{RGB}{255, 165, 0}     %
\definecolor{vulnerableColor}{RGB}{204, 0, 0}         %
\definecolor{exploitPathColor}{RGB}{204, 0, 0}        %

\usepackage{float}
\usepackage[skins]{tcolorbox}
\usepackage{fancybox}
\usepackage{changepage}
\usepackage{blindtext}
\usepackage{enumitem}
\usepackage{multirow}
\usepackage{booktabs}
\usepackage{balance}
\usepackage{subcaption}
\usepackage{url}
\usepackage{soul}
\usepackage{color}
\usepackage{siunitx}
\usepackage{makecell}
\usepackage{colortbl}
\definecolor{deepblue}{rgb}{0,.2,0.6}
\definecolor{deepgreen}{rgb}{0,0.5,0}
\definecolor{deepchampagne}{rgb}{0.98, 0.84, 0.65}
\definecolor{mintgreen}{rgb}{0.6, 1.0, 0.6}
\definecolor{vividviolet}{rgb}{0.62, 0.0, 1.0}
\definecolor{mangotango}{rgb}{1.0, 0.51, 0.26}
\definecolor{dkgreen}{rgb}{0,0.5,0}
\definecolor{dkred}{rgb}{0.5,0,0}
\definecolor{gray}{rgb}{0.5,0.5,0.5}

\usepackage{algorithmic}
\usepackage{graphicx}
\usepackage{textcomp}
\usepackage{xcolor}

\usepackage[acronym]{glossaries}
\glsdisablehyper
\usepackage{longtable}

\usepackage{pifont}
 
\usepackage{listings}
\usepackage{pgfmath}
\lstdefinelanguage{JavaScript}{
  keywords={typeof, new, true, false, catch, function, return, null, catch, switch, var, if, in, while, do, else, case, break, const, await, async, yield},
  keywordstyle=\color{blue},
  ndkeywords={class, export, boolean, throw, implements, import, this},
  ndkeywordstyle=\color{darkgray},
  identifierstyle=\color{black},
  sensitive=false,
  comment=[l]{//},
  morecomment=[s]{/*}{*/},
  commentstyle=\color{olive},
  stringstyle=\color{purple},
  morestring=[b]',
  morestring=[b]"
}
\lstdefinestyle{customjs}{
  language=JavaScript,
  extendedchars=true,
  basicstyle=\small\ttfamily,
  showstringspaces=false,
  showspaces=false,
  numbers=left,
  numberstyle=\footnotesize,
  tabsize=2,
  breaklines=false,
  showtabs=false,
  captionpos=b,
  frame=single,
  framerule=0.2pt,
  morecomment=[l][\annotationcolor\bfseries]{\#},
  framexleftmargin=2em,
  xleftmargin=2.5em,
  xrightmargin=0.5em,
}

\usepackage{hyperref}
\def\BibTeX{{\rm B\kern-.05em{\sc i\kern-.025em b}\kern-.08em
    T\kern-.1667em\lower.7ex\hbox{E}\kern-.125emX}}

\usepackage[normalem]{ulem}

\newcommand{\headerbox}[2]{
\vspace{1mm}  \noindent\fbox{
    \begin{minipage}{0.96\columnwidth}
      \textbf{#1}: #2
    \end{minipage}%
  }%
\vspace{0mm}
}%

\newcommand{\figref}[1]{Figure~\ref{#1}}

\makeatletter
\newsavebox{\@brx}
\providecommand{\llangle}{}\renewcommand{\llangle}[1][]{\savebox{\@brx}{\(\m@th{#1\langle}\)}%
  \mathopen{\copy\@brx\kern-0.5\wd\@brx\usebox{\@brx}}}
\providecommand{\rrangle}{}\renewcommand{\rrangle}[1][]{\savebox{\@brx}{\(\m@th{#1\rangle}\)}%
  \mathclose{\copy\@brx\kern-0.5\wd\@brx\usebox{\@brx}}}
\makeatother

\newacronym{aci}{ACI}{Arbitrary Command Injection}
\newacronym{ace}{ACE}{Arbitrary Code Execution}
\newacronym{llms}{LLMs}{Large Language Models}

\newcommand{\nodemedic}{\textsc{NodeMedic}\xspace}
\newcommand{\nodemedicfine}{\textsc{NodeMedic-FINE}\xspace}
\newcommand{\name}{\textsc{Triage-JS}\xspace}

\newcommand{\nodejs}{{Node.js}\xspace}

\newcommand{\npm}{{npm}\xspace}

\newcommand{\cleartext}[1]{}

\usepackage{listings}
\usepackage{xcolor}
\usepackage{tikz}
\usetikzlibrary{arrows.meta, shapes.geometric, positioning}

\lstdefinelanguage{JavaScript}{
    keywords={function, var, require, if, else, return, try, catch},
    keywordstyle=\color{blue}\bfseries,
    ndkeywords={__set_taint__, exec},
    ndkeywordstyle=\color{orange}\bfseries,
    identifierstyle=\color{black},
    sensitive=false,
    comment=[l]{//},
    morecomment=[s]{/*}{*/},
    commentstyle=\color{gray}\itshape,
    stringstyle=\color{red},
    morestring=[b]',
    morestring=[b]"
}

\usepackage{pifont}
\newcommand{\cmark}{\ding{51}} %
\newcommand{\xmark}{\ding{55}} %

\newcommand{\fionebarvar}[2]{%
  \begin{tikzpicture}[baseline={(0,-2pt)}]
    \pgfmathsetmacro{\mean}{#1}
    \pgfmathparse{#2} \let\var\pgfmathresult
    \pgfmathsetmacro{\stddev}{sqrt(\var)}
    \pgfmathsetmacro{\stddev}{sqrt(\var)}
    \pgfmathsetmacro{\lowraw}{\mean - 2.776 * \stddev}
    \pgfmathsetmacro{\highraw}{\mean + 2.776 * \stddev}

    \pgfmathsetmacro{\minval}{0.85}
    \pgfmathsetmacro{\maxval}{0.95}
    \pgfmathsetmacro{\scalefactor}{.03cm}

    \pgfmathsetmacro{\low}{max(\lowraw,\minval)}
    \pgfmathsetmacro{\high}{min(\highraw,\maxval)}
    \pgfmathsetmacro{\xlow}{(\low - \minval)/(\maxval - \minval) * \scalefactor}
    \pgfmathsetmacro{\xhigh}{(\high - \minval)/(\maxval - \minval) * \scalefactor}

    \pgfmathsetmacro{\xdotraw}{(\mean - \minval)/(\maxval - \minval) * \scalefactor}

    \draw[gray!60, thick] (0, 0) -- (\scalefactor, 0); %

    \ifdim\xhigh pt<0pt
    \else
      \ifdim\xlow pt>\scalefactor pt
      \else
        \draw[black, thick] (\xlow, 0) -- (\xhigh, 0); %
      \fi
    \fi

    \ifdim\xdotraw pt<0pt
    \else
      \ifdim\xdotraw pt>\scalefactor pt
      \else
        \fill[black] (\xdotraw, 0) circle[radius=1pt]; %
      \fi
    \fi
  \end{tikzpicture}%
}

\setlist{topsep=2pt, partopsep=0pt, parsep=1pt, itemsep=1pt}

\setcopyright{cc}
\setcctype{by}
\acmDOI{10.1145/3832783.3837503}
\acmYear{2026}
\copyrightyear{2026}
\acmISBN{979-8-4007-2882-2/2026/10}
\acmConference[ASE '26]{Proceedings of the 41st IEEE/ACM International Conference on Automated Software Engineering}{October 12--16, 2026}{Munich, Germany}
\acmBooktitle{Proceedings of the 41st IEEE/ACM International Conference on Automated Software Engineering (ASE '26), October 12--16, 2026, Munich, Germany}
\acmSubmissionID{ase26main-p1614-p}
\received{2026-03-26}
\received[accepted]{2026-06-18}

\begin{document}

\title{Learning to Triage Vulnerability Reports from Program Analysis: An Empirical Study in Node.js}

\author{Ronghao Ni}
\orcid{0009-0008-8518-5660}
\authornote{Equal contribution.}
\affiliation{%
  \institution{Carnegie Mellon University}
  \city{Pittsburgh}
  \state{Pennsylvania}
  \country{USA}
}
\email{ronghaon@andrew.cmu.edu}

\author{Aidan Z.H. Yang}
\orcid{0009-0003-8808-6426}
\authornotemark[1]
\affiliation{%
  \institution{Athanor AI}
  \city{Seattle}
  \state{Washington}
  \country{USA}
}
\email{aidan@athanorl.com}

\author{Min-Chien Hsu}
\orcid{0009-0004-8165-167X}
\affiliation{%
  \institution{Carnegie Mellon University}
  \city{Pittsburgh}
  \state{Pennsylvania}
  \country{USA}
}
\email{minchieh@andrew.cmu.edu}

\author{Nuno Sabino}
\orcid{0000-0001-6302-477X}
\affiliation{%
  \institution{Carnegie Mellon University}
  \city{Pittsburgh}
  \state{Pennsylvania}
  \country{USA}
}
\email{nsabino@andrew.cmu.edu}

\author{Limin Jia}
\orcid{0000-0002-8160-349X}
\affiliation{%
  \institution{Carnegie Mellon University}
  \city{Pittsburgh}
  \state{Pennsylvania}
  \country{USA}
}
\email{liminjia@andrew.cmu.edu}

\author{Ruben Martins}
\orcid{0000-0003-1525-1382}
\affiliation{%
  \institution{Carnegie Mellon University}
  \city{Pittsburgh}
  \state{Pennsylvania}
  \country{USA}
}
\email{rubenm@andrew.cmu.edu}

\author{Darion Cassel}
\orcid{0000-0002-7898-966X}
\affiliation{%
  \institution{Carnegie Mellon University}
  \city{Berkeley}
  \state{California}
  \country{USA}
}
\email{darion.cassel@gmail.com}

\author{Kevin Cheang}
\orcid{0000-0002-5717-0575}
\affiliation{%
  \institution{UC Berkeley}
  \city{Berkeley}
  \state{California}
  \country{USA}
}
\email{kcheang@berkeley.edu}

\renewcommand{\shortauthors}{R. Ni, A. Z.H. Yang, M.-C. Hsu, N. Sabino,
  L. Jia, R. Martins, D. Cassel, and K. Cheang}

\begin{abstract}

Program analysis tools often produce large volumes of candidate vulnerability reports that require costly manual review, creating a practical challenge: how can security analysts prioritize the reports most likely to be true vulnerabilities? This paper investigates whether machine learning can be applied to prioritizing vulnerabilities reported by program analysis tools. We focus on \nodejs packages and collect a benchmark of 1,883 \nodejs packages, each containing one reported ACE or ACI vulnerability. We evaluate a variety of machine learning approaches, including classical models, graph neural networks (GNNs), large language models (LLMs), and hybrid models that combine GNNs and LLMs, trained on data derived from program analysis tool outputs (\nodemedicfine{} in our case study). The top LLM achieves $F_{1}{=}0.915$, while the best provenance-graph-based method achieves $F_{1}{=}0.904$. On the reports that the upstream tool flags but cannot automatically confirm, at a target of recovering 90\% of the exploitable reports, the leading model eliminates 75\% of the benign reports from manual review. If the best model is tuned to operate at a precision level of 0.8 (i.e., allowing 20\% false positives among all warnings), our approach can report 99.2\% of exploitable taint flows while missing only 0.8\%, demonstrating strong potential for real-world vulnerability triage.

\end{abstract}



\maketitle

\section{Introduction}

Program analysis tools can identify potential security vulnerabilities in software, but often produce a large volume of %
vulnerability reports that require time-intensive manual confirmation by security analysts.
Studies show that developers generally prefer tools 
with low false positives, that
too many false positives are a significant pain point, and that many developers avoid tools that have too many false positives, leading to tool disuse~\cite{christakis2016developers,johnson2013don}. 
One way to reduce false positives is to automatically synthesize proof-of-concept (PoC) exploits that demonstrate that the reported vulnerabilities can %
be exploited by an attacker (e.g.,~\cite{cassel2023nodemedic,nodemedic-fine,kang2023scaling,lekies201325,parameshwaran2015dexterjs,bensalim_talking,garmany_primgen,yaw_hijax,steffens2020pmforce}). 
However, automated synthesis (e.g., \nodemedicfine{}) confirms only a subset of reported flows in practice; it suffers from low recall due to execution-constraint complexity. Even when a PoC exploit exists, analysts still need to review whether the exploit matches the program's threat model. To reduce manual triage effort, we apply machine learning to prioritize reports from existing tools, enabling analysts to focus on those most likely to be exploitable.

\begin{figure}[t]
    \centering
    \resizebox{\columnwidth}{!}{\begin{tikzpicture}[
  >=Latex,
  font=\normalsize,
  node distance=7mm and 10mm,
  box/.style={
    draw=black!65,
    rounded corners=2pt,
    line width=0.45pt,
    minimum height=0.86cm,
    align=center,
    text width=#1,
    inner sep=3pt
  },
  darkbox/.style={box=#1, fill=teal!25, text=black},
  bluebox/.style={box=#1, fill=blue!14},
  greenbox/.style={box=#1, fill=green!13},
  brownbox/.style={box=#1, fill=orange!22},
  graybox/.style={box=#1, fill=gray!18},
  note/.style={font=\small, align=center}
]

\node[bluebox=1.90cm] (analysis) {Program\\Analysis};
\node[bluebox=2.15cm, below=10mm of analysis] (raw) {Vulnerability\\Reports};
\draw[->, line width=0.45pt] (analysis.south) -- node[right, note] {} (raw.north);

\node[darkbox=1.58cm, right=10mm of raw, yshift=15mm] (snippets) {Code\\Snippets};
\node[brownbox=1.58cm, below=4mm of snippets] (prov) {Graph Representations};

\node[darkbox=1.95cm, right=6mm of snippets] (llm) {LLM-Based\\Methods};
\node[brownbox=1.95cm, right=6mm of prov] (gnnml) {GNN/Classical ML\\Methods};
\node[greenbox=1.95cm, below=4mm of gnnml] (hybrid) {Hybrid\\GNN-LLM};

\path ($(snippets.north west)+(-8mm,10mm)$) coordinate (engNW);
\path ($(gnnml.south east)+(8mm,-15mm)$) coordinate (engSE);
\begin{scope}[on background layer]
\draw[draw=black!30, rounded corners=4pt, fill=black!4, line width=0.35pt] (engNW) rectangle (engSE);
\end{scope}
\node[
  fill=teal!25,
  text=black,
  rounded corners=2pt,
  inner xsep=4pt,
  inner ysep=2pt,
  anchor=west,
  font=\normalsize\bfseries
] at ($(engNW)+(2mm,-5mm)$) {ML-BASED TRIAGE ENGINE};

\draw[->, line width=0.45pt] (raw.east) .. controls +(right:8mm) and +(left:10mm) .. node[above=7mm, note, pos=0.5] {} (snippets.west);
\draw[->, line width=0.45pt] (raw.east) -- node[above=-6mm, note, pos=0.56] {} (prov.west);

\draw[->, line width=0.45pt] (snippets.east) -- (llm.west);
\draw[->, line width=0.45pt] (prov.east) -- (gnnml.west);
\draw[->, line width=0.45pt] (snippets.east) -- (hybrid.west);
\draw[->, line width=0.45pt] (prov.east) -- (hybrid.west);

\node[bluebox=1.95cm, right=10mm of gnnml, yshift=10mm] (queue) {Prioritized\\Review Queue};
\node[greenbox=1.80cm, below=10mm of queue] (analyst) {Security Analyst};
\path ($(queue.west)+(-6mm,0)$) coordinate (queue_in);
\draw[->, line width=0.45pt] (llm.east) -- ++(4mm,0) |- (queue_in) -- (queue.west);
\draw[->, line width=0.45pt] (gnnml.east) -- (queue_in) -- (queue.west);
\draw[->, line width=0.45pt] (hybrid.east) -- ++(4mm,0) -| (queue_in) -- (queue.west);
\draw[->, line width=0.45pt] (queue.south) -- node[right, note] {Top-ranked\\reports} (analyst.north);
\end{tikzpicture}}
    \caption{%
    Triage pipeline and evaluated method families.}
    \label{fig:all_exp}
\end{figure}

This work focuses on the \gls{aci} and \gls{ace} vulnerabilities~\cite{cwe_94, cwe_77} in \nodejs %
packages. 
The \nodejs runtime is one of the most popular frameworks for web, desktop, and mobile developers.
Studies have found that the \nodejs ecosystem is full of packages containing vulnerabilities~\cite{duantowards, jang2010empirical, staicu2019empirical, zahan2022weak}. 
\gls{aci} and \gls{ace} are high-severity vulnerabilities that allow an attacker to execute code or commands on the system running the application.
An attacker who achieves \gls{ace} or \gls{aci} controls the host process and can launch other attacks, e.g., directory traversal~\cite{cwe_22}, by extension.
Previous research has introduced automated methods to identify such vulnerabilities~\cite{cassel2023nodemedic, bensalim_talking, nodemedic-fine, yaw_hijax, garmany_primgen}.
Many of these approaches use dynamic taint analysis, %
which attempts to trace the flow of data from attacker-controlled inputs through the program to critical APIs.
Dynamic analyses may report many potentially dangerous flows, but may not always indicate which ones can actually be exploited (i.e., true positives). 
Static analysis techniques often generate even more false positives.

Prior work has applied machine learning to vulnerability detection~\cite{linevul, vuldeepecker, draper, steenhoek2024dataflow, yang2024security}, typically framing it as a classification problem and training models directly on source code or abstract representations such as abstract syntax trees (ASTs) or control flow graphs (CFGs).
However, ACE and ACI vulnerability patterns can involve complex data flow dependencies
which are not readily inferred from source code or textual explanations.
In contrast, dynamic taint analysis tools can cover entire codebases and identify  vulnerabilities across complex data flow patterns.
For instance, vulnerable ACE and ACI flows can be automatically discovered by dynamic taint analysis tools like \nodemedic~\cite{cassel2023nodemedic} or the more recent \nodemedicfine~\cite{nodemedic-fine}, which output {\em taint provenance graphs}
capturing all operations performed on attacker-controlled inputs flowing to the sink~\cite{cassel2023nodemedic,nodemedic-fine}.

This work investigates whether work done by dynamic taint analysis tools can be directly leveraged by ML models to triage \gls{ace}  and \gls{aci} vulnerabilities in \nodejs packages reported by these tools. 
Our insights are the following: 
\textbf{First}, graph-based ML models, including graph neural networks (GNNs) and classical classifiers, can be
configured to serve as vulnerability-report triage tools by performing binary
classification on the provenance graphs.
\textbf{Second}, since recent language models pre-trained on a large corpus of code tokens have a prior understanding of code structures and general vulnerability patterns, fine-tuning these models on a dataset annotated by a dynamic taint analysis tool to include specific vulnerability patterns results in strong vulnerability triage performance.

A summary of our evaluated machine learning triage approaches
is shown in Figure~\ref{fig:all_exp}:
LLM-only, which applies large language models directly to source code; %
ML, which uses classical ML models (Random Forest, XGBoost, Logistic Regression, and SVM) trained on features extracted from provenance graphs generated by the dynamic taint analysis tool \nodemedicfine;
GNN, which uses graph neural networks on the same provenance graphs as in ML;
GNN-LLM, a hybrid model that combines embeddings from GNNs over provenance graphs and pre-trained LLMs over source code.

To evaluate our approaches, we construct a benchmark, \name{}, comprising 1,883 \npm packages for which \nodemedicfine successfully detected potential taint flows.
Designed for report triage, \name{} reflects the deployment setting in which analysts prioritize \nodemedicfine{} outputs from our \npm{} crawl.
We compare against program analysis and synthesis-based approaches for automatically synthesizing PoC exploits~\cite{cassel2023nodemedic,nodemedic-fine,kang2023scaling} as the baselines. 
Our best approach, DS-Distill-Qwen-7B with full fine-tuning and a classification head,
achieves an average F1 score of 0.915. In comparison, \nodemedicfine{} 
alone %
achieves only %
$F_{1}= 0.676$.
For practical vulnerability triage, \textit{on the reports that the upstream tool flags but cannot automatically confirm, at a target of recovering 90\% of the exploitable reports, this model eliminates 75\% of the benign reports from manual review (inspecting 74 of the 119 unconfirmed test reports)}.

In summary, we contribute: (1) \name{}, a manually labeled benchmark of 1,883 \nodejs packages for post-analysis triage of reported ACE/ACI flows; (2) a comparative evaluation of classical ML, GNN, LLM, and GNN+LLM architectures for prioritizing reports from an upstream taint-analysis tool; and (3) evidence that learned triage substantially improves recall on analyzer reports while preserving practical precision, together with an analysis of robustness to simulated vulnerable-to-benign prevalence shifts. The benchmark will be released after responsible disclosure is complete.

\section{Background}
We review \nodejs's threat model, program analysis tools for vulnerability detection and confirmation for \nodejs packages, and existing ML approaches for vulnerability detection.

\paragraph{\nodejs Package Threat Model}
\label{sec:background:threat-model}  
\nodejs is %
built on top of the V8 JavaScript engine. \nodejs developers combine code into {\em packages}, which
can import other packages as \emph{dependencies} to use their public APIs (exported functions).
\nodejs provides powerful sensitive APIs~\cite{eval_fn,new_function_fn,exec_fn,execsync_fn} that 
can dynamically generate and execute code and execute shell commands.

In a real-world attack, 
a \nodejs package that unsafely uses sensitive 
\nodejs APIs is included as a dependency of a victim application as illustrated in \figref{fig:attacker-model}.
An attacker can be any user communicating with the 
victim application.
Attacker-controlled input is passed
from the victim application to a
sensitive API (e.g., \texttt{exec}~\cite{exec_fn}) via the dependency's public API.
We use the same model of the above scenario as prior work~\cite{staicu-SYNODE,nodemedic-fine,cassel2023nodemedic}:
The attacker \emph{directly} passes input to the dependency.
We consider \emph{all} public APIs of the dependency to be the attack surface of the package.

\begin{figure}[t]
  \centering
  \centering
    \resizebox{0.55\linewidth}{!}{%

    \begin{tikzpicture}[
        node_box/.style={
            rectangle,
            rounded corners=6pt,
            fill=#1, %
            text=white,
            font=\bfseries\large\sffamily,
            minimum width=2.6cm,
            minimum height=1.0cm,
            align=center
        },
        std_arrow/.style={
            -{Triangle[width=2.5mm,length=2.5mm]},
            thick,
            color=black!80
        },
        exploit_arrow/.style={
            -{Triangle[width=3mm,length=3mm]},
            very thick,
            dashed,
            color=exploitPathColor %
        }
    ]

        \node[node_box={serverColor}] (server) {Server};

        \node[node_box={intermediateColor}, below left=0.7cm and 0.1cm of server] (dep1) {Dep 1};
        \node[node_box={safeColor}, below right=0.7cm and 0.1cm of server] (dep2) {Dep 2};

        \node[node_box={safeColor}, below=0.7cm of dep1, xshift=-2cm] (dep3) {Dep 3};
        \node[node_box={vulnerableColor}, below=0.7cm of dep1, xshift=2cm] (dep4) {Dep 4}; %
        \node[node_box={safeColor}, below=0.7cm of dep2, xshift=1cm] (dep5) {Dep 5};

        \draw[std_arrow] (server.south) -- (dep1.north);
        \draw[std_arrow] (server.south) -- (dep2.north);
        \draw[std_arrow] (dep1.south) -- (dep3.north);
        \draw[std_arrow] (dep1.south) -- (dep4.north);
        \draw[std_arrow] (dep2.south) -- (dep5.north);

        \draw[exploit_arrow] (server.south) to[bend right=20] (dep1.north);
        \draw[exploit_arrow] (dep1.south) to[bend left=20] (dep4.north);

        \node[above=0.1cm of server, font=\ttfamily\bfseries\huge, text=exploitPathColor] {<exploit>};

        \node[below=0.3cm of dep4, font=\ttfamily\bfseries\Large] {
            exec(\textcolor{vulnerableColor}{<exploit>})
        };

    \end{tikzpicture}
    }
    \vspace{-20pt}
  \caption{\nodejs attacker model. Solid arrows indicate dependency relationships. The dashed red arrows depict the malicious data flow via Dep 1 (orange) to the vulnerable Dep 4 (red), where the payload is executed.}
  \label{fig:attacker-model}
\end{figure}

We use the \texttt{toygrep} utility as a running example (Figure~\ref{fig:toy-provenance-graph}): the tainted \texttt{query} argument is concatenated with a \texttt{grep} command string and passed to an \texttt{exec} sink.
An \gls{aci} occurs when attacker input reaches such a command-execution sink: here, an attacker can append a second command for the shell to execute.
An \gls{ace} is analogous but involves code-evaluation sinks (\texttt{eval}, \texttt{new Function}), where attacker input is interpreted as JavaScript inside the package process.
The underlying injection sinks are pervasive in the \npm ecosystem~\cite{staicu-SYNODE,staicu2019empirical}.
Command injection is also a common class in \nodejs analysis reports; for example, it was the most common vulnerability type reported in ODGen's study of 300K packages~\cite{li2022odgen}.

\paragraph{Dynamic Taint Analysis, Provenance Graphs, and Vulnerability Confirmation}
\label{subsec:provenance-analysis}
Taint analysis specifies and checks policies governing sensitive dataflow in programs. Its goal is to detect flows from attacker-controlled \emph{sources} (e.g., public API inputs) to security-sensitive \emph{sinks} (e.g., command/code execution APIs). Dynamic taint analysis performs this tracking during execution and has been effective for JavaScript security analysis (cf.~\cite{andreasen-Survey}).

In this paper, we use \nodemedicfine~\cite{nodemedic-fine}, which emits a taint \emph{provenance graph}: a history of taint-propagating operations. Nodes represent operations/values and edges represent dataflow dependencies. The vulnerable flow of \texttt{toygrep} yields the provenance structure shown in Figure~\ref{fig:toy-provenance-graph}, where nodes include operations such as \texttt{string.concat} and edges track source-to-sink flow.
\begin{figure}[t!]
    \centering
    \begin{minipage}{0.64\linewidth}  %
        \centering
        \begin{lstlisting}[language=JavaScript, caption=Source code for the toy package \texttt{toygrep}, label=lst:toy_code]
function grep(query) {
    exec("grep " + query);
}
        \end{lstlisting}

        \vspace{0.2cm} %

        \begin{lstlisting}[language=JavaScript, caption=Driver program for the toy package \texttt{toygrep}, label=lst:driver_program]
var PUT = require("toygrep");
var x = "tainted"; // {0:'0'}
__set_taint__(x);
try{PUT.grep(x);}
catch (e) {console.log(e)}
        \end{lstlisting}
    \end{minipage}
    \hfill
    \begin{minipage}{0.34\linewidth}  %
        \centering
        \begin{tikzpicture}[
            scale=0.84, %
            node distance=1.2cm, %
            every node/.style={draw, align=center, font=\small, ellipse, minimum height=0.8cm, minimum width=2.5cm, transform shape},
            normal/.style={draw=black},
            taintededge/.style={-{Stealth}, thick, red},
            untaintededge/.style={-{Stealth}, thick, black},
            taintflow/.style={-{Stealth}, thick, red}
        ]

        \node[normal] (n1) {Untainted \\ {[String: 'grep ']}};
        \node[normal, below=0.7cm of n1] (n2) {call:grep \\ 'tainted'};
        \node[normal, below=0.6cm of n2] (n3) {string.concat \\ 'grep tainted'};
        \node[normal, below=0.6cm of n3] (n4) {call:exec \\ 'grep tainted'};

        \draw[taintededge] (n2) -- (n3);  %
        \draw[taintflow] (n3) -- (n4);  %

        \draw[taintflow] ([xshift=-1.2cm, yshift=0.4cm]n2.north) to[out=0, in=90] (n2.north);

        \draw[untaintededge] (n1.south east) to[out=-50, in=50] (n3.north east);

        \end{tikzpicture}
    \end{minipage}
    \caption{Provenance graph generated from the \texttt{toygrep} package on the right. Ovals are nodes. Black edges are untainted, red edges are tainted flows.
Listing \ref{lst:toy_code} is the source code. Listing \ref{lst:driver_program} is the driver program. }
    \label{fig:toy-provenance-graph}
\end{figure}

\nodemedicfine~\cite{nodemedic-fine} uses the provenance graph generated by the dynamic taint analysis to synthesize a proof-of-concept exploit
by mapping operations in the provenance graph to an SMTLIB constraint satisfaction formula, solving 
for package inputs that can deploy a test payload at the sink.
The package is then executed with the synthesized inputs; successful payload 
execution (e.g., creating a target file) confirms the vulnerability.
Each node stores operation and location metadata used by downstream triage models; Table~\ref{tab:node_attributes} summarizes these attributes.

\begin{table}[t]
\centering
\caption{Attributes of nodes in provenance graphs.\label{tab:node_attributes}}
\resizebox{0.85\columnwidth}{!}{%
\begin{tabular}{@{}p{2cm}p{5.5cm}@{}}
\toprule
\textbf{Attribute}          & \textbf{Description}                       \\ \midrule
\textbf{Operation}         & Type of operation, e.g., \texttt{call} or \texttt{Untainted}. \\ \midrule
\textbf{Value}             & Input provided to the operation.           \\ \midrule
\textbf{File Path}         & File where the operation occurs.           \\ \midrule
\textbf{Position}          & Start and end line/column in the file.     \\ \midrule
\textbf{Tainted Status}    & Whether the data is tainted or untainted.  \\ \midrule
\textbf{Flows From}        & Predecessor nodes in the data flow.        \\ \midrule
\textbf{Sink Type}         & \texttt{eval} (ACE) or \texttt{exec} (ACI). \\ \bottomrule
\end{tabular}%
}
\end{table}

Explode.js~\cite{marques2025automated} and FAST~\cite{kang2023scaling} use symbolic execution to collect constraints, which can be solved to generate \emph{potential} exploits. 
FAST generates candidate exploits without confirming them; and thus can report false positives. 
Explode.js additionally synthesizes complex drivers that can consist of a chain of calls, starting from a public API to eventually reach the function that contains the sink.

\paragraph{Machine Learning Vulnerability Detection}
Classical approaches, such as Support Vector Machines (SVMs) and Random Forests, have been widely applied to program vulnerability detection~\cite{perl2015vccfinder, draper, hanif2021rise}.
These methods typically rely on a variety of features for classification, including program traces, call graphs, literals, variables, data types, operators, and statements.
However, these methods lack the ability to \textit{simulate} program execution.
More modern architectures, such as deep neural networks and graph neural networks, have been applied~\cite{Lin2020SoftwareVD, chakraborty2021deep, steenhoek2024dataflow} to bridge this gap by fitting to highly nonlinear patterns and simulating dataflow,
while large language models have shown promise in reasoning about program semantics~\cite{lu2024grace, huang-chang-2023}.
Prior work has explored a variety of usage modes: in zero-shot settings, LLMs are prompted to make predictions directly without task-specific tuning; in few-shot settings, they are provided with a handful of labeled examples to guide inference; and in fine-tuned settings, the model weights are updated on domain-specific data to specialize behavior.

Our work explores how these %
methods 
can be repurposed to assist triage rather than detection. Specifically, we use taint provenance graphs (generated by \nodemedicfine~\cite{nodemedic-fine} in our evaluation) and relevant source code as inputs to ML models, enabling them to predict which reported flows are truly exploitable and thus prioritize analyst attention more effectively.

\section{Dataset and Methodology}

\lstdefinestyle{figsnippet}{
  language=JavaScript,
  numbers=none,
  basicstyle=\ttfamily\fontsize{7}{7.6}\selectfont,
  xleftmargin=0pt,
  xrightmargin=0pt,
  aboveskip=2pt,
  belowskip=2pt,
  frame=none,
  backgroundcolor=\color{black!3},
  columns=fullflexible
}

In this section, we first explain how \name{}, our dedicated \nodejs taint flow triage benchmark, is constructed. Then we detail our approaches for taint flow triage, as shown in Figure~\ref{fig:all_exp}. Throughout the paper, a \nodemedicfine{} report refers to a candidate \gls{ace}/\gls{aci} taint flow flagged in a package and represented by a taint-provenance graph. For triage, we pair this graph with the surrounding source-code context.
We use \nodemedicfine as the underlying analyzer; transfer to another analyzer would require compatible artifacts and re-validation.
LLM-based methods read snippets around reported sinks, while graph-based and hybrid methods consume provenance graphs. 
Concretely, we evaluate three method families. LLM-based methods classify vulnerability reports from code context using zero-shot prompting or supervised fine-tuning variants (Section~\ref{sec:llm-methods}). Graph-based methods (GNN and classical ML) operate on structured graph representations of reported flows (Section~\ref{sec:gnn-ml-methods}). The hybrid GNN-LLM method concatenates graph and code embeddings and trains a joint classifier to combine complementary signals (Section~\ref{sec:gnn-llm-methods}).

\subsection{\name{: A Triage Benchmark}}
\label{subsec:datasets}
\name{} contains 1,883 \npm packages for which \nodemedicfine successfully detected potential \gls{ace} or \gls{aci} vulnerabilities and generated valid provenance graphs. %

To provide context on data curation, we started from an initial pool of 33,011 \npm packages obtained from the authors of \nodemedicfine. These were %
collected from the \npm registry and pre-filtered to include packages with at least one weekly download and at least one call to a supported sink. Running \nodemedicfine on this pool produced 2,051 packages with %
taint-flow reports.
We excluded 168 packages due to malformed or incomplete analysis artifacts (e.g., missing sinks in output graphs or invalid file/line metadata), yielding the final 1,883-package benchmark. %

Of the 1,883 potentially vulnerable \npm packages, 644 were confirmed by \nodemedicfine{}'s exploit synthesis component as containing exploitable vulnerabilities (i.e., a working proof-of-concept exploit was generated automatically). The remaining 1,239 packages were examined manually.
 Annotation was performed by two of the authors, both with software-security and program-analysis expertise. Each package was first reviewed independently: the reviewers inspected the reported taint flow, relevant code context, and the potential for attacker-controlled input to reach security-sensitive sinks. A package was labeled exploitable only when a concrete attack appeared feasible; disagreements and cases requiring judgment were resolved through adjudication between the two annotators to produce a consistent final label.

\begin{figure}[b]
\begin{minipage}[t]{0.48\linewidth}
\textbf{\footnotesize Direct code injection}
\begin{lstlisting}[style=figsnippet]
obj = eval('(' + input + ')');
\end{lstlisting}
\end{minipage}
\hfill
\raisebox{-0.42in}{\tikz{\draw[densely dashed,gray!65,line width=0.35pt] (0,0) -- (0,0.62in);}}
\hfill
\begin{minipage}[t]{0.48\linewidth}
\textbf{\footnotesize Incomplete sanitizer}
\begin{lstlisting}[style=figsnippet]
input = input.replace(/"/, '');
exec(`bin "${input}"`);
\end{lstlisting}
\end{minipage}

\setlength{\abovecaptionskip}{3pt}
\caption{Representative exploitable flows. The first splices attacker input
directly into a code string (\gls{ace}); the second attempts an incomplete
sanitizer, removing only the first quote, so the command injection
(\gls{aci}) still succeeds.}
\label{fig:vuln-patterns}
\end{figure}

\begin{figure}[b]
\begin{minipage}[t]{0.27\linewidth}
\makebox[\linewidth][l]{\textbf{\footnotesize Validation guard}}
\begin{lstlisting}[style=figsnippet]
if (!re.test(x))
  return;
exec(`cmd ${x}`);
\end{lstlisting}
\end{minipage}
\hfill
\raisebox{-0.42in}{\tikz{\draw[densely dashed,gray!65,line width=0.35pt] (0,0) -- (0,0.62in);}}
\hfill
\begin{minipage}[t]{0.285\linewidth}
\makebox[\linewidth][l]{\textbf{\footnotesize Over-tainting}}
\begin{lstlisting}[style=figsnippet]
id = parseInt(x);
if (isNaN(id))
  return;
exec(`cmd ${id}`);
\end{lstlisting}
\end{minipage}
\hfill
\raisebox{-0.42in}{\tikz{\draw[densely dashed,gray!65,line width=0.35pt] (0,0) -- (0,0.62in);}}
\hfill
\begin{minipage}[t]{0.365\linewidth}
\makebox[\linewidth][l]{\textbf{\footnotesize Sink-family}}
\begin{lstlisting}[style=figsnippet]
spawn('bin', ['sub', x]);
\end{lstlisting}
\end{minipage}
\setlength{\abovecaptionskip}{3pt}
\caption{The three false-alarm modes: validation guard, over-tainting, and
sink-family semantics. \nodemedicfine{} reports these flows
because it does not model the validation effect, keeps taint through an
imprecise numeric coercion, or treats separate \texttt{spawn} arguments like a
shell-interpreted command string.}
\label{fig:fa-patterns}
\end{figure}

\paragraph{Benchmark characterization.}
The exploitable flows are dominated by direct injection: in about half of them, attacker-controlled input is concatenated directly into the command or code string that reaches the sink (Figure~\ref{fig:vuln-patterns}). Others are subtler; for instance, a developer attempts to sanitize the input with an incomplete guard. %
Among the false alarms, three patterns recur (Figure~\ref{fig:fa-patterns}). The first is \emph{input validation and sanitization}: attacker data reaches the sink, but it first passes a check whose effect \nodemedicfine{} does not model, so the value stays tainted even though dangerous characters have been rejected. The second is \emph{over-approximated taint propagation}: to avoid missing flows, \nodemedicfine{} keeps a value tainted through operations it cannot track precisely (marked \texttt{imprecise}), even when the operation constrains or removes the attacker's control. The third, and most common, is \emph{sink-family semantics}: \nodemedicfine{} treats all supported sinks alike, but sinks such as \texttt{spawn} pass attacker-controlled data as separate arguments rather than concatenating it into a shell-interpreted command string, so shell metacharacters do not create a second command.

Since ML methods require parts of the data for training and validation, we randomly divided the 1,883-package dataset into train/validation/test splits with an 8:1:1 ratio (1,506/188/189 packages) (see Table \ref{tab:dataset-overview} for details).
For models requiring training, we use the train split for optimization and the validation split for model selection; all methods are reported on the same test split.

\begin{table}[t]
\centering
\setlength{\belowcaptionskip}{3pt}
\caption{The dataset splits used in the evaluation, showing the total number of packages in each split (\textit{train}, \textit{validate}, and \textit{test}), along with the number of vulnerable packages in each split, categorized into ACE and ACI. Numbers in parentheses indicate the count of vulnerable packages within each split.}
\label{tab:dataset-overview}
\resizebox{0.88\columnwidth}{!}{%
\begin{tabular}{lccc}
\toprule
\textbf{Split} & \textbf{Total (Vuln.)} & \textbf{ACE (Vuln.)} & \textbf{ACI (Vuln.)} \\
\midrule
train & 1,506 (989) & 255 (176) & 1,251 (813) \\
validate & 188 (124) & 36 (23) & 152 (101) \\
test & 189 (137) & 38 (29) & 151 (108) \\
\midrule
total & 1,883 (1,250) & 329 (228) & 1,554 (1,022) \\
\bottomrule
\end{tabular}
}
\end{table}

\subsection{LLM-based Methods}
\label{sec:llm-methods}
Common practices in using large language models (LLMs) for classification tasks typically fall into two categories:
    \textbf{(1) Zero-shot and few-shot prompting}, where the model is provided with code snippets along with carefully designed natural language prompts to identify security risks. This approach benefits from LLMs' generalization ability but often struggles with nuanced vulnerabilities that require deeper program understanding.
    \textbf{(2) Fine-tuning}, where the model is trained on labeled datasets to learn domain-specific patterns. Fine-tuning can significantly improve detection accuracy but comes with high computational costs and data collection challenges. Beyond full fine-tuning, there are also lightweight fine-tuning methods, such as LoRA fine-tuning \cite{lora, qlora} and fine-tuning on only selected layers. Commonly, a language model can either be fine-tuned on a text generation objective or used as a classifier by attaching a classification head on top of the pre-trained model. 
    We adopt the latter approach, treating the models as classifiers whose outputs are logits used for prediction.

To systematically evaluate LLMs in taint flow triage, we assess several models under different settings, including zero-shot classification, linear probing, LoRA fine-tuning \cite{lora, qlora}, and full fine-tuning. We exclude few-shot learning in this work because the large size of potentially vulnerable package code snippets makes it difficult to fit multiple samples into a reasonable context window.
The code snippet given to LLMs is extracted from the file containing the reported sink and truncated to 1,024 tokens around the sink and nearby reported data-flow context. This tool-guided truncation strategy keeps the input focused on the relevant code; our observations confirm that most vulnerable logic is local to the sink and surrounding code in the dataset.

\subsubsection{Zero-Shot Classification}

In this %
setting, we use an auto-regressive generation head that enables the LLM to generate a textual response indicating whether a given JavaScript package contains %
vulnerabilities. %
Zero-shot classification relies entirely on the LLM's pre-trained knowledge and ability to generate a relevant answer token by token in an auto-regressive manner.
For this method, we prompt the LLM with the following query:

\vspace{-3pt}
\begin{tcolorbox}[colback=gray!10, colframe=black, arc=3mm, boxrule=0.8pt, top=1pt, bottom=1pt]
\small
\textbf{User:} Our dynamic analysis tool identified a taint flow in a Node.js package, suggesting a potential vulnerability related to either arbitrary code execution (CWE-094) or arbitrary command injection (CWE-078). While the tool attempts to confirm vulnerabilities by generating exploits, this approach may miss some cases. I hope you can assist with triaging and classification by predicting whether the vulnerability is exploitable.
\\\\
I have extracted relevant parts of the code from the file containing the sink, along with surrounding lines for context. After reasoning about the snippet, please output ``Yes'' if you believe it contains an exploitable vulnerability, or ``No'' if you believe it is not exploitable.\\

    \texttt{\{code snippet here\}}
\end{tcolorbox}

For models that run locally, we disable sampling during generation to ensure deterministic results. For cloud-based API models, we mitigate run-to-run variability by repeating each evaluation five times and reporting mean and variance.
The outputs are filtered based on the presence of ``Yes'' which is considered vulnerable, while all other cases, including those that generate neither ``Yes'' nor ``No'' are considered non-vulnerable. We use zero-shot prompting only as a reference setting; our main reproducible results use fine-tuned classifiers that read class logits directly.
Structured outputs and few-shot prompting could make prompting-based triage more robust; we leave these directions to future work.

\subsubsection{Linear Probing}

Instead of using a generation head, we attach a classification head on top of the base LLMs and train the head for the taint flow triage task while keeping the base LLMs frozen.
In our setup, the classification head is just a single linear layer (an affine linear transformation).
The classification head takes the embedding from the last non-padding position of the output from the last attention layer as input and produces two logits, representing the two classes (vulnerable or non-vulnerable). We use the cross-entropy loss function to train the classification head with weights that correspond to the class imbalance in the dataset.

\subsubsection{LoRA Fine-Tuning}

LoRA (Low-Rank Adaptation) fine-tuning \cite{lora, qlora} offers a lightweight approach to adapting LLMs without updating all parameters. Instead of modifying the entire model, LoRA injects low-rank adapters into selected layers of the model. These adapters are small, low-rank matrices that are learned during fine-tuning. Similar to the previous method, we attach a classification head on top of the base LLMs and fine-tune the model for the taint flow triage task. However, in this case, we only update the low-rank adapters and the classification head, while the base LLMs remain frozen. 

\subsubsection{Full Fine-Tuning}

In this setting, we update all parameters of the LLM and the classification head using our labeled benchmark \name{}, applying the same cross-entropy loss function as previously described.
During the fine-tuning of all the aforementioned LLM-based methods, all frozen parameters are stored in 4-bit NormalFloat (NF4) precision for memory efficiency \cite{qlora}, while the trainable parameters are in 16-bit BrainFloat (BF16) precision.

\subsection{GNN and Classical ML Methods}
\label{sec:gnn-ml-methods}
The GNN and ML methods utilize the provenance graphs (Section \ref{subsec:provenance-analysis}) created by \nodemedicfine's
taint provenance tracking %
in both the training and inference pipelines. The operation, tainted status of the arguments, and sink type of each node are used as inputs. %
Additionally, the vulnerability type is included as an input for the entire graph. The 100 most common operations in our dataset (as described in Section \ref{subsec:datasets}) are assigned class numbers \texttt{0} to \texttt{99}. Class \texttt{100} is designated for less frequent operations, while class \texttt{101} is used for empty or missing operation attributes in the provenance graphs. Tainted statuses are encoded as class \texttt{0} for \texttt{False} (untainted), class \texttt{1} for \texttt{True} (tainted), and class \texttt{2} for missing attributes. Sink types are represented with class \texttt{0} for \texttt{spawn}, class \texttt{1} for \texttt{exec}, class \texttt{2} for \texttt{Function}, and class \texttt{3} for \texttt{eval}. Vulnerability types are encoded as class \texttt{0} for ACE and class \texttt{1} for ACI vulnerabilities.

Each attribute is represented as a one-hot vector, where the corresponding class has a value of \texttt{1}, and all other classes have a value of \texttt{0}. The four one-hot vectors are then concatenated to form the embedding for a single node in the graph. Together, the graph's node connections and the embeddings of its nodes make up the complete representation of the graph. %

\subsubsection{GNN}

The GNN component %
starts with a Gated Graph Sequence Neural Network (GGNN) \cite{li2015gated}, which is a specialized type of neural network designed to learn from graph-structured data by capturing dependencies and relationships between nodes. The GGNN works by iteratively passing messages along edges, enabling each node to gather information from its neighbors and update its representation based on the graph’s structure and features. In the final step, the learned abstract node embeddings are combined into a graph-level representation using Global Attention Pooling~\cite{li2015gated}, resulting in the final graph embedding. The graph embedding is then fed into a classification head to predict vulnerability.\sloppy %

\begin{table}[t]
    \centering
    \caption{Model Comparison Across %
    Experiment Settings.}
    \label{tab:llm_experiments}
    \resizebox{\columnwidth}{!}{%
    \begin{tabular}{lcccc}
        \toprule
        \textbf{LLM Model} & \textbf{Zero-Shot} & \textbf{Linear Probing} & \textbf{LoRA FT} & \textbf{Full FT}\\
        \midrule
        Claude Sonnet 4.5 & \cmark & \xmark & \xmark & \xmark \\
        OpenAI GPT-5 & \cmark & \xmark & \xmark & \xmark \\
        OpenAI o4-mini-high & \cmark & \xmark & \xmark & \xmark \\
        DeepSeek-R1-0528 & \cmark & \xmark & \xmark & \xmark \\
        Gemini 2.5 Pro & \cmark & \xmark & \xmark & \xmark \\
        OpenAI GPT-4.1 & \cmark & \xmark & \xmark & \xmark \\
        \midrule
        DeepSeek-R1-Distill-Qwen-14B & \cmark & \cmark & \cmark & \xmark \\
        DeepSeek-R1-Distill-Llama-8B & \cmark & \cmark & \cmark & \cmark \\
        DeepSeek-R1-Distill-Qwen-7B & \cmark & \cmark & \cmark & \cmark \\
        Llama-3.1-8B(-Instruct) & \cmark & \cmark & \cmark & \cmark \\
        Qwen2.5-Coder-14B(-Instruct) & \cmark & \cmark & \cmark & \xmark \\
        Qwen2.5-Coder-7B(-Instruct) & \cmark & \cmark & \cmark & \cmark \\
        \bottomrule
    \end{tabular}
    }
\end{table}

\subsubsection{ML}

Classical ML methods use only the graph's node embeddings and discard the edges. The embeddings of all nodes are first fed into a pooling layer to create a fixed-size embedding vector that represents the entire graph, regardless of the number of nodes. The pooled embedding is then passed through machine learning classifiers to predict the vulnerability.

\subsection{GNN-LLM Hybrid Methods}
\label{sec:gnn-llm-methods}
We evaluate a hybrid approach that combines \nodemedicfine{}'s vulnerability-path reports with graph neural networks (GNNs) and large language models (LLMs). This method, referred to as GNN-LLM, aims to integrate the strengths of both program analysis and LLMs to enhance taint flow triage. %
This hybrid approach employs two parallel encoding branches. The \emph{graph branch} processes the provenance graph through a GGNN to learn structural dependencies between taint sources, sinks, and intermediate operations, producing a graph-level embedding. The \emph{code branch} encodes the tool-guided truncated code snippet through an LLM backbone to capture semantic context and code-level details, producing a code embedding from the final layer. These two embeddings are concatenated to form a joint representation that fuses structural (graph) and semantic (code) information. This concatenated vector is then passed to a single classification head, which is trained to predict whether the input contains a vulnerability.
The full model is trained end-to-end, with gradients backpropagating through both the GNN and LLM components simultaneously, allowing them to adapt and specialize their representations for the triage task.

\section{Experimental Setup}
We first outline the model selection and implementation (Section \ref{subsec:model_selection}) used in our evaluation. Next, we discuss the evaluation metrics (Section \ref{subsec:experiment_metrics}) and the system configuration (Section \ref{subsec:system_config}).

\subsection{Model Selection and Implementation}\label{subsec:model_selection}
We directly report results of \nodemedicfine's exploit synthesis obtained from \nodemedicfine's authors. We downloaded FAST~\cite{kang2023scaling} and ran it
against \name{} with the \texttt{-X} flag enabled to turn on exploit generation. 
Additionally, for FAST, we used the \texttt{-t} flag to specify vulnerability types as \texttt{os\_command} and \texttt{code\_exec}, which correspond to ACI and ACE vulnerabilities, respectively.

For classical ML methods, we evaluate %
logistic regression, support vector machine (SVM), random forest, and XGBoost, which are trained on provenance graphs generated by \nodemedicfine. For logistic regression, SVM, and random forest experiments, we used classes from the \texttt{scikit-learn} library. The \texttt{xgboost} package was used for XGBoost experiments. For these machine learning baseline models, default hyperparameters were applied. The GNN method is implemented using the \texttt{torch-geometric} library \cite{Fey/Lenssen/2019}. The model is trained with a learning rate of 0.001, a batch size of 64, and a weight decay rate of 0.1. The GNN model is trained for 150 epochs, with early stopping based on validation F1 scores.

An overview of the LLM models and their experiment settings is in Table \ref{tab:llm_experiments}.
We experiment with local language models and their pretrained variants: DeepSeek-R1-Distill-Qwen-14B, DeepSeek-R1-Distill-Llama-8B, DeepSeek-R1-Distill-Qwen-7B~\cite{deepseek2025deepseek}, Llama-3.1-8B~\cite{dubey2024llama}, Qwen2.5-Coder-14B, and Qwen2.5-Coder-7B~\cite{qwen2.5}.
We use the publicly available implementations and parameters from Hugging Face. For Qwen and Llama models, we use the instruct-tuned versions for zero-shot evaluation, and the corresponding base models for finetuning with a classification head.

Specifically, we evaluate zero-shot performance using Claude Sonnet 4.5, OpenAI GPT-5, OpenAI o4-mini-high, DeepSeek R1-0528~\cite{deepseek2025deepseek}, Gemini 2.5 Pro~\cite{comanici2025gemini}, and OpenAI GPT-4.1~\cite{achiam2023gpt}, accessed via OpenRouter. These models represent widely used commercial model families and deployment tiers. To mitigate API run-to-run variability, each API model follows the same five-repetition protocol as the local models, and we report the mean and variance; provider-side model updates remain outside our control.

For LoRA finetuning, we use a rank of 128 and an alpha of 64 for all experiments. A study on LoRA's hyperparameters is presented in Section~\ref{subsec:llm-tuning}. All LLM methods that require fine-tuning are trained with a batch size of 2 per device, a learning rate of 1e-5, and a weight decay rate of 0.01. The training is conducted for 3 epochs, with early stopping based on validation F1 scores.
In brief, the main trained configurations use AdamW with learning rate $1\times10^{-5}$, batch size 2 per GPU, and up to 3 epochs for LLM tuning; LoRA uses rank 128 and alpha 64; and the GGNN uses learning rate $1\times10^{-3}$, batch size 64, and up to 150 epochs.

\subsection{Evaluation Metrics}\label{subsec:experiment_metrics}

\nodemedicfine, FAST, and zero-shot LLM models are evaluated directly on the test dataset, while other models that require training are trained on the training dataset and validated on the validation dataset during the training process. The best model from training is then evaluated on the test dataset.
To assess the performance of each method, we follow prior work on vulnerability detection~\cite{steenhoek2024dataflow, ivdetect, linevul, yang2024security}
 and employ the following metrics: 
$\text{F1} = \frac{TP}{TP + 0.5 \cdot (FP + FN)}$, 
$\text{Precision} = \frac{TP}{TP + FP}$, 
$\text{Recall} = \frac{TP}{TP + FN}$, and 
$\text{Accuracy} = \frac{TP + TN}{TP + TN + FP + FN}$
 (where TN is true negative, TP is true positive, FP is false positive, and FN is false negative).
For all methods, we report the average of the metrics across five runs, each with a different random seed (2025 through 2029). The average and variance of the metrics are reported in Section \ref{sec:results}.
For precision-recall curves shown in Section~\ref{subsec:tradeoffs}, we also compute average precision (AP) across five runs at each recall level. The average precision is defined as $\mathrm{AP} = \sum_{n=1}^{N} (R_n - R_{n-1}) \cdot P_n$
, where $P_n$ and $R_n$ are the precision and recall at threshold index $n$.

\subsection{System Configuration}\label{subsec:system_config}

Evaluations that require only CPUs are conducted on a computing cluster. Each task runs with a 36-hour timeout %
in a virtually isolated environment with a 2-core CPU, which is part of an AMD EPYC 7742 processor, and 16 GB of RAM.
 For experiments that require GPUs, except for those where LLMs are undergoing full fine-tuning, each task is executed in a virtual environment with one NVIDIA H100 (80GB) GPU, two Intel Xeon 8480C PCIe Gen5 CPUs
, and 2 TB of RAM. For LLMs that require full fine-tuning, we use a computing cluster with two NVIDIA H100 (80GB) GPUs.

\begin{table*}[t!]\centering
    \caption{F1 Score (F1), Precision (Prec), Recall (Rec), and Accuracy (Acc). Higher metrics indicate better performance. For models with multiple configurations or usage variations, results are reported for the setup with the highest F1 score. 
    All metrics are reported as the mean and variance over five runs with different random seeds. 
    A dash (–) denotes an undefined metric due to division by zero or a single-setup model.
    \textit{Bold} indicates the best F1 score within each group; \underline{underline} indicates the best F1 score overall.
    F1 confidence intervals are visualized using horizontal bars centered at the mean, computed as $\bar{x} \pm 2.776 \cdot \text{SD}$ (95\% confidence with $n=5$, Student's $t$ distribution). Intervals outside the fixed axis range (0.85–0.95) are clipped for display.
    }\label{tab:main_results}
    \renewcommand{\arraystretch}{1.0}
    \resizebox{0.97\textwidth}{!}{%
    \begin{tabular}{llrrrrrrrrr}
    \toprule
    \multirow{2}{*}{\textbf{Model}} 
    & \multirow{2}{*}{\textbf{Best Setup}} 
    & \multicolumn{3}{c}{F1} 
    & \multicolumn{2}{c}{Prec} 
    & \multicolumn{2}{c}{Rec} 
    & \multicolumn{2}{c}{Acc} \\
    \cmidrule(lr){3-5} \cmidrule(lr){6-7} \cmidrule(lr){8-9} \cmidrule(lr){10-11}
    & & Mean & Var & CI Bar & Mean & Var & Mean & Var & Mean & Var \\
    \midrule
    \textbf{Random ($P_{vuln}=1/2$)} & — & 0.592 & 0.00e+00 & \fionebarvar{0.592}{0.00e+00} & 0.725 & 0.00e+00 & 0.500 & 0.00e+00 & 0.500 & 0.00e+00 \\
    \textbf{Random ($P_{vuln}=989/1506$)} & — & 0.689 & 0.00e+00 & \fionebarvar{0.689}{0.00e+00} & 0.725 & 0.00e+00 & 0.657 & 0.00e+00 & 0.571 & 0.00e+00 \\
    \textbf{Random ($P_{vuln}=1$)} & — & \textbf{0.841} & 0.00e+00 & \fionebarvar{0.841}{0.00e+00} & 0.725 & 0.00e+00 & 1.000 & 0.00e+00 & 0.725 & 0.00e+00 \\
    \textbf{Random ($P_{vuln}=0$)} & — & 0.000 & 0.00e+00 & \fionebarvar{0.000}{0.00e+00} & -- & -- & 0.000 & 0.00e+00 & 0.275 & 0.00e+00 \\
    \midrule
    \textbf{\nodemedicfine} & — & \textbf{0.676} & 0.00e+00 & \fionebarvar{0.676}{0.00e+00} & 1.000 & 0.00e+00 & 0.511 & 0.00e+00 & 0.646 & 0.00e+00 \\
    \textbf{FAST} & — & 0.647 & 0.00e+00 & \fionebarvar{0.647}{0.00e+00} & 0.957 & 0.00e+00 & 0.489 & 0.00e+00 & 0.614 & 0.00e+00 \\
    \midrule
    \textbf{GNN} & — & 0.886 & 1.71e-04 & \fionebarvar{0.886}{1.71e-04} & 0.914 & 6.99e-05 & 0.858 & 3.36e-04 & 0.839 & 3.02e-04 \\
    \textbf{Random Forest} & Avg Pooling & 0.900 & 8.37e-06 & \fionebarvar{0.900}{0.0837e-04} & 0.917 & 1.99e-07 & 0.883 & 2.66e-05 & 0.857 & 1.40e-05 \\
    \textbf{XGBoost} & Avg Pooling & \textbf{0.904} & 0.00e+00 & \fionebarvar{0.904}{0.00e+00} & 0.917 & 0.00e+00 & 0.891 & 0.00e+00 & 0.862 & 0.00e+00 \\
    \textbf{Logistic Regression} & Max Pooling & 0.892 & 0.00e+00 & \fionebarvar{0.892}{0.00e+00} & 0.909 & 0.00e+00 & 0.876 & 0.00e+00 & 0.847 & 0.00e+00 \\
    \textbf{SVM} & Max Pooling & 0.898 & 0.00e+00 & \fionebarvar{0.898}{0.00e+00} & 0.898 & 0.00e+00 & 0.898 & 0.00e+00 & 0.852 & 0.00e+00 \\
    \midrule
    \textbf{Claude Sonnet 4.5} & Zero-shot & 0.858 & 5.72e-05 & \fionebarvar{0.858}{0.572e-04} & 0.819 & 9.24e-05 & 0.902 & 3.62e-04 & 0.784 & 1.04e-04 \\
    \textbf{OpenAI GPT-5} & Zero-shot & 0.848 & 9.67e-05 & \fionebarvar{0.848}{0.967e-04} & 0.867 & 1.03e-04 & 0.829 & 2.29e-04 & 0.784 & 1.74e-04 \\
    \textbf{OpenAI o4-mini-high} & Zero-shot & 0.805 & 7.49e-05 & \fionebarvar{0.805}{0.749e-04} & 0.889 & 5.64e-05 & 0.736 & 2.77e-04 & 0.742 & 7.53e-05 \\
    \textbf{DeepSeek R1-0528} & Zero-shot & 0.857 & 5.70e-05 & \fionebarvar{0.857}{0.570e-04} & 0.824 & 1.04e-04 & 0.893 & 3.36e-04 & 0.784 & 1.03e-04 \\
    \textbf{Gemini 2.5 Pro} & Zero-shot  & 0.851 & 1.85e-04 & \fionebarvar{0.851}{1.85e-04} & 0.769 & 3.95e-04 & 0.953 & 1.49e-04 & 0.758 & 6.08e-04 \\
    \textbf{OpenAI GPT-4.1} & Zero-shot  & 0.858 & 9.20e-06 & \fionebarvar{0.858}{0.092e-04} & 0.778 & 7.45e-06 & 0.956 & 2.66e-05 & 0.770 & 2.25e-05 \\
    \textbf{DeepSeek-R1-Distill-Qwen-14B} & LoRA FT & 0.866 & 1.29e-04 & \fionebarvar{0.866}{1.29e-04} & 0.861 & 2.90e-04 & 0.872 & 2.29e-04 & 0.804 & 2.94e-04 \\
    \textbf{DeepSeek-R1-Distill-Llama-8B} & LoRA FT & 0.889 & 2.70e-04 & \fionebarvar{0.889}{2.70e-04} & 0.878 & 1.93e-04 & 0.901 & 5.75e-04 & 0.837 & 5.37e-04 \\
    \textbf{DeepSeek-R1-Distill-Qwen-7B} & Full FT & \underline{\textbf{0.915}} & 1.72e-04 & \fionebarvar{0.915}{1.72e-04} & 0.900 & 1.90e-04 & 0.931 & 2.56e-04 & 0.875 & 3.72e-04 \\
    \textbf{Llama-3.1-8B(-Instruct)} & LoRA FT & 0.909 & 9.46e-05 & \fionebarvar{0.909}{0.946e-04} & 0.888 & 6.67e-05 & 0.930 & 2.29e-04 & 0.865 & 1.90e-04 \\
    \textbf{Qwen2.5-Coder-14B(-Instruct)} & LoRA FT & 0.863 & 1.18e-04 & \fionebarvar{0.863}{1.18e-04} & 0.846 & 8.87e-05 & 0.880 & 4.95e-04 & 0.797 & 2.04e-04 \\
    \textbf{Qwen2.5-Coder-7B(-Instruct)} & Full+GNN FT & 0.902 & 6.26e-05 & \fionebarvar{0.902}{0.626e-04} & 0.882 & 4.27e-04 & 0.923 & 3.89e-04 & 0.854 & 1.62e-04 \\
    \bottomrule
    \end{tabular}
    }
\end{table*}

\section{Evaluation}
\label{sec:results}

Our experiments address the following research questions:
\begin{itemize}[left=-5pt, label={}]
    \setlength{\itemsep}{0pt}
    \item \textbf{RQ1: Effectiveness and Robustness.} How effective are classical ML, GNN, and LLM-based methods at triaging reported taint flows, and how robust are they in more benign-heavy settings?
    \item \textbf{RQ2: Comparison of Model Architectures.} How do graph-based and LLM-based methods compare, do they agree on predictions, and which local LLM usage strategies are most effective?
    \item \textbf{RQ3: Method Efficiency.} How do machine learning methods compare to traditional taint-analysis approaches in training overhead and latency?
    \item \textbf{RQ4: Practical Triage Value.} How much review effort can learned triage reduce at fixed precision/recall targets, and how do cost constraints affect model choice?
\end{itemize}

\subsection{RQ1: Effectiveness and Robustness}
\label{subsection:effectiveness}
\subsubsection{Primary-Benchmark Effectiveness}
Table~\ref{tab:main_results} shows that learned triage methods substantially improve over synthesis-only confirmation. \nodemedicfine achieves high precision but low recall (F1 0.676), reflecting %
the high failure rate of exploit synthesis.

As calibration baselines, we include random classifiers that label each package as vulnerable independently at random with a fixed probability $P_{vuln}$; Table~\ref{tab:main_results} reports $P_{vuln}\in\{0,\,1/2,\,989/1506,\,1\}$, where $989/1506$ is the vulnerable fraction of the training split. The highest-F1 random baseline, $P_{vuln}=1$ (label everything vulnerable), attains F1 0.841, but offers no triage value because it does not reduce analyst workload. In contrast, learned models improve both ranking quality and thresholded decision quality.

Among graph-driven methods, GNN reaches F1 0.886 and XGBoost reaches F1 0.904, indicating strong triage performance from provenance-graph features. Fine-tuned local LLMs perform best overall: DS-Distill-Qwen-7B (Full FT) reaches F1 0.915, with Llama-3.1-8B (LoRA FT) close behind at F1 0.909.

\begin{figure}[t]
  \centering
  \setlength{\belowcaptionskip}{0pt}%
  \begin{subfigure}[t]{0.8\linewidth}
    \centering
    \includegraphics[width=0.8\linewidth]{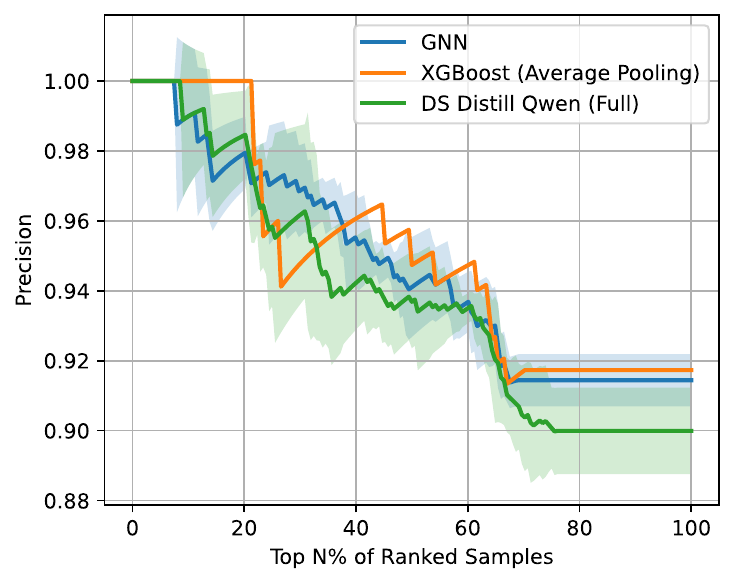}
    \caption{Precision at top N\%.}
  \end{subfigure}
  \vspace{0.2em}
  \begin{subfigure}[t]{0.8\linewidth}
    \centering
    \includegraphics[width=0.8\linewidth]{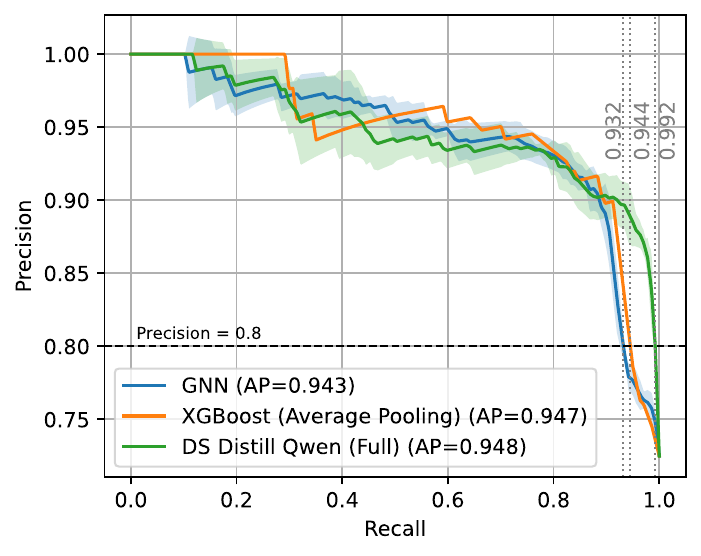}
    \caption{Precision-recall curve.}
    \label{fig:triage_dual_pr}
  \end{subfigure}
  \caption{Ranking and threshold behavior for GNN, XGBoost, and DS-Distill-Qwen-7B.}
  \label{fig:triage_dual}
\end{figure}

Figure~\ref{fig:triage_dual} shows practical ranking behavior. The best models keep high precision in top-ranked outputs and maintain strong precision-recall characteristics under threshold tuning.
This ranking-oriented analysis is operationally important because developers frequently inspect only a small prefix of reported findings. Section~\ref{subsec:usage} returns to this behavior when discussing manual review effort.

\subsubsection{Robustness to Simulated Vulnerable-to-Benign Ratios}
\label{subsec:ratio}

The primary benchmark, \name{}, reflects the empirical prevalence an analyst sees among
reports emitted by \nodemedicfine{} on the sink-filtered \npm{} registry crawl used to construct the
benchmark (Section~\ref{subsec:datasets}). Among its flagged flows, $1{,}250$ are
exploitable and $633$ are false alarms, a ratio of about $2{:}1$. We extend our study to
more benign-heavy ratios because deployment conditions can differ: other
analyzers may have higher false-alarm rates, and a different package population
may contain many more benign flows. 

\begin{table}[t]
\centering
\setlength{\belowcaptionskip}{3pt}
\caption{F$_1$ robustness to the vulnerable-to-benign ratio (5-seed means).
The \textit{Original} column reports results on the original test distribution:
$137$ vulnerable and $52$ benign reports ($\approx 2.63{:}1$; Table~\ref{tab:main_results}).
The ``re-weighted'' rows keep each model fixed
and re-weight the test set to each ratio. The ``retrained'' rows retrain at each
ratio and test at that same ratio. \textbf{Random} ($P_{vuln}{=}1$) is the
all-vulnerable baseline, the highest-F$_1$ random baseline possible.}
\label{tab:ratio_robustness}
\begingroup
\setlength{\tabcolsep}{4pt}
\resizebox{0.9\columnwidth}{!}{%
\begin{tabular}{@{}llcccc@{}}
\toprule
Model & Setting & Original & $1{:}1$ & $1{:}2$ & $1{:}4$ \\
\midrule
\multirow{2}{*}{XGBoost}
  & re-weighted & 0.90 & 0.85 & 0.77 & 0.65 \\
  & retrained   & 0.90 & 0.83 & 0.74 & 0.64 \\
\addlinespace[1pt]
\multirow{2}{*}{GNN}
  & re-weighted & 0.89 & 0.83 & 0.75 & 0.63 \\
  & retrained   & 0.89 & 0.83 & 0.74 & 0.67 \\
\addlinespace[1pt]
\multirow{2}{*}{DS-Qwen-7B (Full)}
  & re-weighted & 0.92 & 0.85 & 0.75 & 0.62 \\
  & retrained   & 0.92 & 0.82 & 0.73 & 0.59 \\
\addlinespace[1pt]
\multirow{2}{*}{Hybrid (DS-Qwen-7B+GNN)}
  & re-weighted & 0.91 & 0.83 & 0.74 & 0.59 \\
  & retrained   & 0.91 & 0.83 & 0.74 & 0.57 \\
\midrule
\multicolumn{2}{@{}l}{Random ($P_{vuln}{=}1$, best)} & 0.84 & 0.67 & 0.50 & 0.33 \\
\bottomrule
\end{tabular}%
}
\endgroup
\end{table}

We simulate a more benign-heavy ratio with two related checks, both shown in
Table~\ref{tab:ratio_robustness}. In the first check (the ``re-weighted'' rows)
we keep each trained model fixed and recompute precision and F$_1$ at more
benign-heavy ratios, up to $1{:}4$ vulnerable to benign, by counting each
held-out benign package more heavily. The
positives are untouched, so recall does not move. At
a fixed threshold the model flags a fixed fraction of benign packages as
vulnerable, so more benign packages mean more false positives while the true
positives stay the same, which lowers precision ($\text{TP}/(\text{TP}+\text{FP})$)
and therefore F$_1$. At $1{:}4$, the fixed models achieve F$_1$ between $0.59$ and $0.65$, retaining a $0.26$--$0.32$ F$_1$-point advantage over the all-vulnerable random baseline of $0.33$ (last row).
In the second check (the ``retrained'' rows) we retrain each model at each
benign-heavy ratio and test it at that same ratio, to test whether training on
the benign-heavy data helps. 
F$_1$ remains close to the fixed-model result, ranging from $0.57$ to $0.67$ at $1{:}4$ and retaining a $0.24$--$0.34$ F$_1$-point advantage over the same baseline. This is
expected because the precision loss comes from the higher proportion of benign
packages in the test set, which retraining does not change.

\headerbox{RQ1 Summary}{In the primary evaluation, fine-tuned local LLMs perform best (F$_1$ 0.915), while XGBoost and GNN reach F$_1$ 0.904 and 0.886, respectively, compared with 0.676 for \nodemedicfine{}'s synthesis-only baseline. At the most benign-heavy simulated ratio ($1{:}4$), all four methods remain $0.24$--$0.34$ F$_1$ points above the best random baseline.}

\subsection{RQ2: Comparison of Model Architectures}
\label{subsec:gnn-llm}
\begin{table}[b]
\centering
\footnotesize
\caption{Performance comparison between fully fine-tuned LLMs with and without GNN components. Each cell shows the average and variance on separate lines.}
\resizebox{0.9\columnwidth}{!}{%
\begin{tabular}{lcccc}
\toprule
\textbf{Model} & \textbf{F1} & \textbf{Precision} & \textbf{Recall} & \textbf{Accuracy} \\
\midrule
\multicolumn{5}{l}{\textbf{DeepSeek-R1-Distill-Llama-8B}} \\
\addlinespace[1pt]
Full         & 0.6837      & 0.6627      & 0.7153      & 0.6794 \\
             & (±1.47e-01) & (±1.41e-01) & (±1.68e-01) & (±5.39e-02) \\
Full + GNN   & 0.5512      & 0.6920      & 0.5226      & 0.5820 \\
             & (±1.48e-01) & (±1.53e-01) & (±1.68e-01) & (±5.44e-02) \\
\midrule
\multicolumn{5}{l}{\textbf{DeepSeek-R1-Distill-Qwen-7B}} \\
\addlinespace[1pt]
Full         & 0.9153      & 0.8999      & 0.9314      & 0.8751 \\
             & (±1.72e-04) & (±1.90e-04) & (±2.56e-04) & (±3.72e-04) \\
Full + GNN   & 0.9116      & 0.8901      & 0.9343      & 0.8688 \\
             & (±1.25e-04) & (±3.74e-05) & (±3.46e-04) & (±2.44e-04) \\
\midrule
\multicolumn{5}{l}{\textbf{Llama-3.1-8B}} \\
\addlinespace[1pt]
Full         & 0.5046      & 0.6672      & 0.5299      & 0.5386 \\
             & (±1.61e-01) & (±1.53e-01) & (±2.32e-01) & (±4.76e-02) \\
Full + GNN   & 0.8312      & 0.8492      & 0.8161      & 0.7608 \\
             & (±5.58e-04) & (±3.62e-04) & (±2.91e-03) & (±6.49e-04) \\
\midrule
\multicolumn{5}{l}{\textbf{Qwen2.5-Coder-7B}} \\
\addlinespace[1pt]
Full         & 0.8962      & 0.8931      & 0.9007      & 0.8497 \\
             & (±6.17e-04) & (±5.91e-05) & (±2.68e-03) & (±9.88e-04) \\
Full + GNN   & 0.9016      & 0.8821      & 0.9226      & 0.8540 \\
             & (±6.26e-05) & (±4.27e-04) & (±3.89e-04) & (±1.62e-04) \\
\bottomrule
\end{tabular}
}
\label{tab:llm_gnn_comparison}
\end{table}

To assess whether provenance-graph information complements code-based LLM representations, Table~\ref{tab:llm_gnn_comparison} compares fully fine-tuned LLMs with and without GNN embeddings. The hybrid models show limited or inconsistent gains, raising the possibility that the component models make similar decisions. We test this possibility through prediction-agreement analysis next.

\subsubsection{Prediction Agreement Analysis}
\begin{figure}[t]
  \centering
  \includegraphics[width=\linewidth]{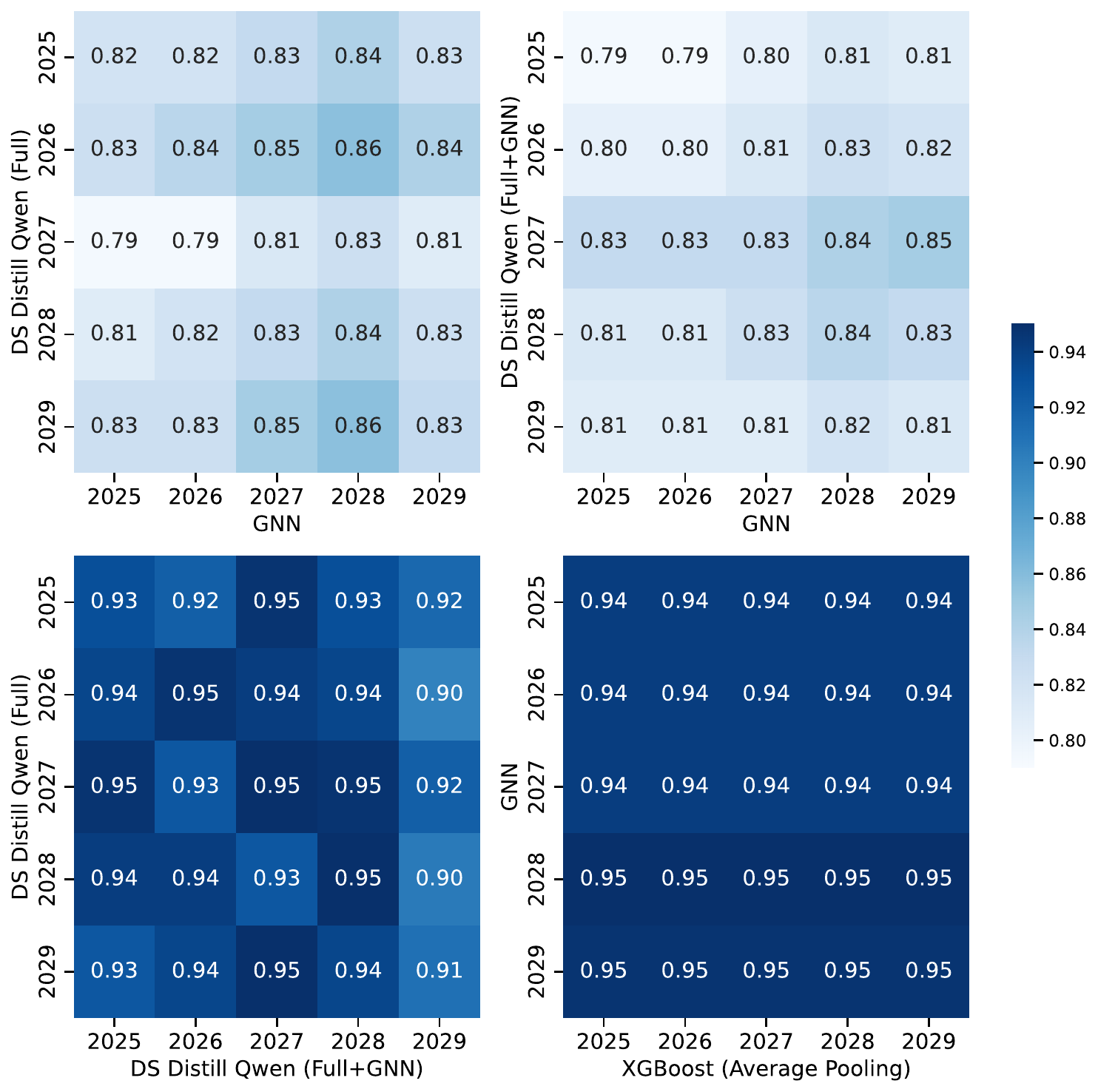}
  \caption{Cohen's Kappa for model-pair prediction agreement.}
  \label{fig:kappa_plot}
\end{figure}

To test this hypothesis, we use Cohen's Kappa~\cite{cohen1960coefficient} on test predictions to compare four model pairs: DS-Distill-Qwen-7B (Full) vs. GNN, DS-Distill-Qwen-7B (Full+GNN) vs. GNN, DS-Distill-Qwen-7B (Full) vs. DS-Distill-Qwen-7B (Full+GNN), and GNN vs. XGBoost. We observe strong agreement between DS-Distill-Qwen-7B (Full) and GNN (average $\kappa=0.83$). This agreement is consistent with the limited gains from fusion: they often make the same decisions, suggesting overlap in their predictive signals (Figure~\ref{fig:kappa_plot}).

\subsubsection{API-Level Pattern Insight via SHAP}
\begin{figure}[t]
  \centering
  \includegraphics[width=\linewidth]{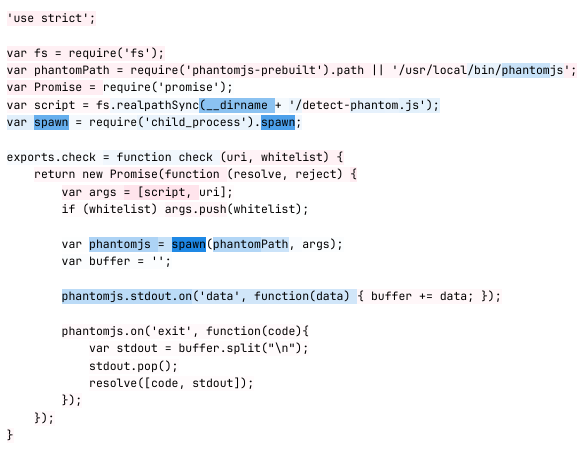}
  \caption{SHAP example for a correctly classified non-vulnerable package.}
  \label{fig:shapley_spawn}
\end{figure}

SHAP (SHapley Additive exPlanations)~\cite{NIPS2017_7062} attributes a model's prediction to its inputs using Shapley values from cooperative game theory: each input token receives its average marginal contribution to the prediction over feature coalitions, so tokens with positive values push the model toward the predicted class and tokens with negative values push against it. Figure~\ref{fig:shapley_spawn} shows an example SHAP analysis. SHAP analysis of DS-Distill-Qwen-7B highlights an actionable API-level pattern: \textit{spawn} often contributes to safe predictions, while \textit{exec} is more frequently associated with vulnerable outcomes. This aligns with API semantics: \textit{spawn} separates command and argument vectors, whereas \textit{exec} interprets a shell string directly. In our dataset, this structural distinction is a useful safety cue that LLMs capture more reliably than pure graph topology.

\subsubsection{Learning False-Alarm Patterns}
\label{subsec:famodes}
The models do not only learn the sink type. They also learn the three false-alarm patterns in Section~\ref{subsec:datasets}. To check this, we fit a logistic regression probe on the pooled node operations, which are the features used by the classical models. The largest weight toward the vulnerable label is \texttt{precise:string.concat}, which means attacker input is directly added to a command or code string. The largest weights toward the benign label are \texttt{spawn}, validation-check operations, and the imprecise coercion operations (\texttt{imprecise:stringify}, \texttt{imprecise:String}). These correspond to safer \texttt{spawn} usage, explicit input checking, and cases where \nodemedicfine{} propagates taint through an imprecise operation even though the final value is not injectable. 
Overall, this suggests that the operation distributions in a flow, together with the sink type, provide useful signals for vulnerability prediction.

\subsubsection{Shared Errors}
\label{subsec:blindspots}
The false-alarm patterns above explain why the models can reject many benign reports, but a small set of hard cases remains difficult for nearly every model family. Treating each architecture as a family (the GNN, the classical classifiers, and the fine-tuned and probed LLMs, nine in all), $114$ of the $189$ test packages are classified correctly by every family, and only $7$ are misclassified by nearly all of them in every seed. These shared misclassifications arise because taint reaching a dangerous sink is a strong signal, but it does not always determine whether the final runtime value is injectable. Most of the shared misclassifications are false alarms: tainted data reaches \texttt{exec}, \texttt{spawn}, or \texttt{Function}, but the relevant operation constrains the runtime value enough that attacker-controlled input cannot change the executed command or code. Distinguishing these cases requires concrete value semantics, such as the exact sanitized value or whether the sink parses the value as command or code syntax.

\subsubsection{LLM Usage Strategies}
\label{subsec:llm-tuning}
For reproducible local deployment, we focus on DS-Distill-Qwen-7B and Llama-3.1-8B. DS-Distill-Qwen-7B achieves the best overall result with full fine-tuning (F1 0.915). Llama-3.1-8B performs strongly with LoRA fine-tuning (F1 0.909) and is more stable than full fine-tuning in our runs, suggesting a practical trade-off between training cost and robustness.

To provide a fuller view of local-model strategy sensitivity, Table~\ref{tab:llm_f1_split} reports F1 across zero-shot prompting, linear probing, LoRA fine-tuning, and full fine-tuning. The cloud-API zero-shot baselines are already reported in Table~\ref{tab:main_results}.

\begin{table}[!htp]\centering
\small
\caption{F1 scores (with variances) for local large language models (LLMs) under different usage strategies: zero-shot inference, linear probing (LP), LoRA-based fine-tuning (LoRA FT), and full model fine-tuning (Full FT). Cloud-API zero-shot baselines are reported in Table~\ref{tab:main_results}. A dash (--) indicates that the method is not evaluated for the model.}
\label{tab:llm_f1_split}
\begin{tabular}{lcccc}
\toprule
\textbf{Zero-shot} & \textbf{LP} & \textbf{LoRA FT} & \textbf{Full FT} \\
\midrule
\multicolumn{4}{l}{\textbf{DeepSeek-R1-Distill-Qwen-14B}} \\
0.816 \scriptsize{(±7.4e-05)} & 0.788 \scriptsize{(±3.2e-04)} & 0.866 \scriptsize{(±1.3e-04)} & -- \\[0.2ex]
\multicolumn{4}{l}{\textbf{DeepSeek-R1-Distill-Llama-8B}} \\
0.793 \scriptsize{(±8.1e-05)} & 0.773 \scriptsize{(±8.7e-04)} & 0.889 \scriptsize{(±2.7e-04)} & 0.684 \scriptsize{(±1.5e-01)} \\[0.2ex]
\multicolumn{4}{l}{\textbf{DeepSeek-R1-Distill-Qwen-7B}} \\
0.723 \scriptsize{(±8.4e-04)} & 0.779 \scriptsize{(±5.4e-04)} & 0.845 \scriptsize{(±2.8e-04)} & 0.915 \scriptsize{(±1.7e-04)} \\[0.2ex]
\multicolumn{4}{l}{\textbf{Llama-3.1-8B(-Instruct)}} \\
0.303 \scriptsize{(±1.0e-03)} & 0.794 \scriptsize{(±3.6e-04)} & 0.909 \scriptsize{(±9.5e-05)} & 0.505 \scriptsize{(±1.6e-01)} \\[0.2ex]
\multicolumn{4}{l}{\textbf{Qwen2.5-Coder-14B(-Instruct)}} \\
0.679 \scriptsize{(±2.7e-04)} & 0.729 \scriptsize{(±1.6e-03)} & 0.863 \scriptsize{(±1.2e-04)} & -- \\[0.2ex]
\multicolumn{4}{l}{\textbf{Qwen2.5-Coder-7B(-Instruct)}} \\
0.526 \scriptsize{(±4.2e-04)} & 0.744 \scriptsize{(±5.3e-04)} & 0.859 \scriptsize{(±3.6e-04)} & 0.896 \scriptsize{(±6.2e-04)} \\
\bottomrule
\end{tabular}%
\end{table}

\paragraph{Zero-shot vs. tuned models.}
For local models, zero-shot performance varies substantially across model families and generally underperforms supervised tuning in this task. Distilled variants typically outperform their non-distilled counterparts under the same usage mode, suggesting that distillation contributes useful task-relevant inductive bias for exploitability triage.

\begin{figure}[t]
  \centering
  \includegraphics[width=0.85\linewidth]{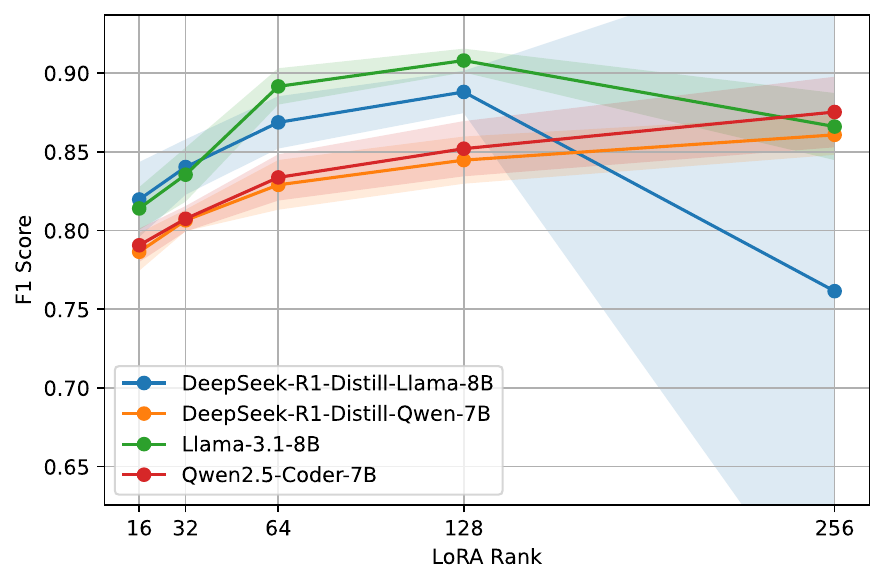}
  \caption{F1 by LoRA rank for representative local models.}
  \label{fig:lora_plot}
\end{figure}

\paragraph{LoRA vs. full fine-tuning.}
The best strategy is model-family dependent. DS-Distill-Qwen-7B achieves its strongest performance with full fine-tuning, while Llama-family models often show better stability under LoRA. These observations align with the practical recommendation to use LoRA when training stability and resource efficiency are priorities, and full fine-tuning when maximizing raw recall is the primary objective.
Figure~\ref{fig:lora_plot} visualizes this trend across representative local models and LoRA ranks.

\headerbox{RQ2 Summary}{Graph and LLM predictors exhibit high agreement ($\kappa=0.83$), and adding GNN embeddings to strong LLM baselines yields limited gains. The false-alarm-mode analysis shows that the models learn signals beyond sink type, while the remaining shared errors involve value semantics not fully captured by the operation sequence.
}

\begin{table}[!htp]\centering
    \small
    \caption{Training and inference times for different models and methods. Inference is measured per sample in a batched setting when batching is supported. Model saving and loading times are not included. The learned-method rows report additional triage cost after \nodemedicfine{} has identified candidate reports and sink locations; the \nodemedicfine{} row shows the upstream detection time.}
    \label{tab: time_overhead}
    \begin{tabular}{lrr}\toprule
    \multirow{2}{*}{Model} & \multicolumn{2}{c}{Computation Time} \\ 
    \cmidrule(lr){2-3}
    & Training & Inference \\ \midrule
    \textbf{\nodemedicfine}  &0min &0.79s \\
    \textbf{FAST} &0min &31.6s \\
    \midrule
    \textbf{GNN} &30s &0.3ms \\
    \textbf{Random Forest} &0.5s &32$\mu$s \\
    \textbf{XGBoost} &0.3s &7$\mu$s \\
    \textbf{Logistic Regression} &1.2s &4$\mu$s \\
    \textbf{SVM} &6.9s &320$\mu$s \\
    \midrule
    \textbf{DeepSeek-R1-Distill-Qwen-14B} &\cellcolor[HTML]{A8A8A8} &\cellcolor[HTML]{A8A8A8} \\
    --zero-shot &0min &15.37s \\
    \textbf{DeepSeek-R1-Distill-Llama-8B} &\cellcolor[HTML]{A8A8A8} &\cellcolor[HTML]{A8A8A8} \\
    --zero-shot &0min &22.72s \\
    \textbf{DeepSeek-R1-Distill-Qwen-7B} &\cellcolor[HTML]{A8A8A8} &\cellcolor[HTML]{A8A8A8} \\
    --zero-shot &0min &5.04s \\
    \textbf{Llama-3.1-8B(-Instruct)} &\cellcolor[HTML]{A8A8A8} &\cellcolor[HTML]{A8A8A8} \\
    --zero-shot &0min &0.48s \\
    --linear-probing &11mins &0.11s \\
    --lora-ft &34mins &0.11s \\
    --full-ft &70mins &0.10s \\
    \textbf{Qwen2.5-Coder-14B(-Instruct)} &\cellcolor[HTML]{A8A8A8} &\cellcolor[HTML]{A8A8A8} \\
    --zero-shot &0min &0.50s \\
    --linear-probing &19mins &0.21s \\
    --lora-ft &58mins &0.21s \\
    \textbf{Qwen2.5-Coder-7B(-Instruct)} &\cellcolor[HTML]{A8A8A8} &\cellcolor[HTML]{A8A8A8} \\
    --zero-shot &0min &0.25s \\
    --linear-probing &10mins &0.10s \\
    --lora-ft &30mins &0.10s \\
    --full-ft &60mins &0.06s \\
    \bottomrule
    \end{tabular}
\end{table}

\subsection{RQ3: Method Efficiency}
\label{subsec:efficiency}
Table~\ref{tab: time_overhead} shows that fine-tuned LLMs cost more to train than the classical and GNN models but still infer quickly. The table reports fine-tuned timings once per architecture: the DeepSeek-R1 distilled models share the architectures of their Qwen2.5 and Llama-3.1 base models, so their fine-tuned training and inference costs match the corresponding base-model rows. The best model, DS-Distill-Qwen-7B with full fine-tuning, thus runs at about $60$\,ms per package, fast enough for automated screening during development. The classical classifiers are cheaper by a wide margin. XGBoost fits in under a second and adds only ${\sim}7\,\mu$s of CPU inference per sample. These models need no GPU at any stage, whereas the LLMs need GPU hardware for both training and inference. Since XGBoost reaches F1 0.904 against the best LLM's 0.915 (Table~\ref{tab:main_results}), XGBoost or Random Forest are better fits when hardware or training budget is tight, while the fine-tuned LLM is preferable when maximizing recall at a fixed precision is the priority (RQ4).

Table~\ref{tab: time_overhead} separates upstream detection from learned triage. All learned methods start after \nodemedicfine{} identifies a candidate report and sink location: the LLM rows measure inference on extracted code, while the classical, GNN, and hybrid rows also use provenance features from the report. End-to-end latency therefore adds the detection time in the \nodemedicfine{} row, about $0.79$\,s per package (about $20$ minutes for the full training set).

\headerbox{RQ3 Summary}{Learned triage shifts cost to offline training but keeps online screening fast (${\approx}60$\,ms per package for the best fine-tuned LLM, microseconds for the classical classifiers on top of the shared feature extraction), making it practical for automated development workflows and complementary to synthesis-based confirmation.}

\subsection{RQ4: Practical Triage Value}
\label{subsec:tradeoffs}%
\label{subsec:usage}
Beyond aggregate accuracy, we evaluate three practical uses of learned triage: choosing a precision--recall operating point, ranking reports that exploit synthesis cannot confirm, and selecting a model that matches a team's resource budget and recall needs.

\subsubsection{Operating Points and Review-Load Reduction}
We compute precision-recall curves using standard sklearn evaluation with interpolation onto a
shared recall axis across seeds (Figure~\ref{fig:triage_dual}\subref{fig:triage_dual_pr}). Based
on prior user studies~\cite{christakis2016developers}, developers generally prefer tools with
precision at or above $0.8$. At a precision operating point of $0.800$, \mbox{DS-Distill-Qwen-7B}
reaches recall $0.992$, versus $0.944$ for XGBoost and $0.932$ for GNN;
this operating point retains nearly all exploitable flows while sharply reducing
benign-package review load.

\subsubsection{Triaging Reports Not Confirmed by Synthesis}
The value of learned triage is clearest where exploit synthesis falls short. Across all $1{,}883$ reports in the primary benchmark, including its train, validation, and test splits, \nodemedicfine{} auto-confirms $644$ reports through exploit synthesis. The remaining $1{,}239$ reports lack a synthesized proof-of-concept and require manual inspection; $606$ are exploitable and $633$ are false alarms.
Because ML models are evaluated on the primary test split, we estimate manual-review savings on the test-split portion of this unconfirmed queue. This test queue contains $119$ reports ($67$ exploitable, $52$ false alarms), all of which are unconfirmed by \nodemedicfine{}. Learned triage does not replace manual validation; it orders these reports so analysts inspect the most likely exploitable cases first. To recover $90\%$ of the exploitable reports, the LLM ranking requires $74$ inspections versus an expected $107$ under random ordering, reducing the review queue by $31\%$ and eliminating $75\%$ of the false alarms from review.
This $31\%$ is close to the maximum achievable: because the queue is already $56\%$ exploitable, even an ideal ranker that surfaces every vulnerability first needs $61$ inspections (a $43\%$ reduction). The LLM therefore captures $72\%$ of the achievable review-load reduction.

\subsubsection{Model Selection by Usage Pattern}
Which model to deploy depends on the team's priority. For cheap, high-volume screening,
XGBoost is a good fit when CPU-only deployment and training cost dominate, while its F$_1$ stays within about a point of the best model. When the priority is to catch reports that exploit synthesis could not confirm, the fine-tuned LLM provides higher recall at the cost of GPU resources. 
In short, the classical models better match cost- and throughput-limited deployments, while the LLM better matches settings where missing an exploitable unconfirmed report is more costly than GPU inference.

\headerbox{RQ4 Summary}{Learned triage is most useful for prioritizing reports that exploit synthesis cannot confirm. XGBoost is a low-cost option for high-throughput screening, while the fine-tuned LLM is preferable when maximizing recall on unconfirmed exploitable reports matters most.}

\section{Threats to Validity}
\textbf{Labeling accuracy.} Manual labeling was required for 1,239 packages not auto-confirmed by synthesis; despite two-annotator adjudication (Section~\ref{subsec:datasets}), residual human error is possible.
Although reviewers used conservative criteria and exploitable-flow context, borderline cases can still be challenging.

\textbf{Single upstream analyzer.} All samples come from one dynamic taint-analysis tool, so the reports' sink coverage and false-alarm profile reflect \nodemedicfine{}'s heuristics, and the models may inherit these biases. Triage also sees only emitted reports, so upstream false negatives remain out of scope by design. Transfer to other languages, vulnerability classes, or upstream analyzers may require retraining and re-validation.

\section{Related Work}

\paragraph{Taint analysis of Node.js packages.}
Several tools perform dynamic or static taint analysis of Node.js 
packages~\cite{cassel2023nodemedic,karim-PlatformIndependent,nodemedic-fine,gauthier-AFFOGATO,nielsen-Nodest,kang2023scaling}.
\nodemedic~\cite{cassel2023nodemedic} generates a provenance graph as a witness of an uncovered tainted flow, and \nodemedicfine~\cite{nodemedic-fine} extends it to find and confirm more flows.
In contrast to these analyzers, our work treats \nodemedicfine's reports as inputs to post-analysis triage; the same framing could apply to any analyzer that emits a graph-structured witness of a reported flow.

\paragraph{ML-based vulnerability detection.}
Static and dynamic techniques have been widely studied for vulnerability detection~\cite{sokMemory,secVulSurvey,fuzzingSurvey}, including code-similarity methods~\cite{vuddy, vulpecker, vuldeepecker} and pattern-based methods~\cite{neuhaus2007predicting, chucky}, which still rely on expert confirmation or handcrafted patterns. Recent ML systems reduce this dependence using AST or data-flow representations with GNNs~\cite{devign,ivdetect,linevd,steenhoek2024dataflow}, language-model attention~\cite{linevul}, or JavaScript-specific taint features~\cite{melicher_lighweight_2021,she_neutaint_2020}. These systems target vulnerability detection, localization, or taint-sink influence; our task is exploitability classification for vulnerability reports already produced by an analyzer.

\paragraph{LLM-based vulnerability analysis.}
LLMs have been increasingly leveraged to support vulnerability detection and triage, spanning fine-grained localization, static-analysis enhancement, and false-positive reduction.
Recent systems fine-tune LLMs for vulnerability localization~\cite{llmao,yang2024security}, use LLM guidance to improve static analysis~\cite{li2024enhancing}, or apply LLM reasoning to warning triage and false-positive pruning~\cite{wen2024automatically,mohajer2024effectiveness}.
Unlike these systems, our work evaluates traditional ML and recent LLMs as exploitability classifiers for vulnerabilities reported by taint-analysis tools in Node.js packages (\nodemedicfine{} in our case study).

\section{Conclusion}
\label{sec:conclusion}
This paper presents a learning-based triage approach for vulnerability reports from program analysis, together with \name{}, a benchmark of 1,883 \nodejs packages for ACE/ACI post-analysis triage. The approach prioritizes analyzer outputs so analysts can focus on the most consequential reports. Both graph-based models and fine-tuned language models learn useful triage signals, with fine-tuned LLMs providing the strongest standalone results.

Our results support learned triage as an optional post-processing step and show that combining analysis-generated evidence with data-driven ranking improves tool usability. Future work can extend this paradigm to other vulnerability classes and ecosystems, explore richer graph--code fusion (e.g., cross-attention), and co-design analyzer outputs with ML models to improve performance, portability, calibration, and robustness.

\section*{Data Availability Statement}
We follow a coordinated vulnerability disclosure process (i.e., responsible disclosure)~\cite{coordinated_vuln_disclosure}.
We are currently reporting the confirmed vulnerabilities included in our dataset to the affected developers.
The full benchmark dataset will be made publicly available after the responsible disclosure process is complete. The code and supporting artifact are available at \url{https://doi.org/10.5281/zenodo.21387402}. The artifact also includes tables listing all hyperparameters used in our experiments, along with the full results table. The artifact and future updates are also maintained at \url{https://github.com/RogerNi/triage-js}.

\begin{acks}
This work was supported by the Future Enterprise Security Initiative at Carnegie Mellon CyLab (FutureEnterprise@CyLab) and by the National Science Foundation under Award No.~2504353.
\end{acks}

\balance
\bibliographystyle{ACM-Reference-Format}
\bibliography{references}

\end{document}